\documentclass{article}

\usepackage{arxiv}

\usepackage[utf8]{inputenc} 
\usepackage[T1]{fontenc}    
\usepackage{hyperref}       
\usepackage{url}            
\usepackage{booktabs}       
\usepackage{amsfonts}       
\usepackage{nicefrac}       
\usepackage{microtype}      
\usepackage{lipsum}
\usepackage{graphicx}
\usepackage[english]{babel}

\title{Electrocatalyst discovery through text mining and multi-objective optimization}

\author{
Lei Zhang\\
Interdisciplinary Centre for Advanced Materials Simulation\\
Ruhr-University Bochum\\
Universit\"atsstra\ss e 150\\
44780 Bochum, Germany\\
\texttt{lei.zhang-w2i@rub.de}\\
	\And
	Markus Stricker\\
Interdisciplinary Centre for Advanced Materials Simulation\\
Ruhr-University Bochum\\
Universit\"atsstra\ss e 150\\
44780 Bochum, Germany\\
\texttt{markus.stricker@rub.de}
}

\begin{document}
\maketitle

\begin{abstract}
The discovery and optimization of high-performance materials is the basis for advancing energy conversion technologies.
To understand composition-property relationships, all available data sources should be leveraged: experimental results, predictions from simulations, and latent knowledge from scientific texts.
Among these three, text-based data sources are still not used to their full potential.
We present an approach combining text mining, Word2Vec representations of materials and properties, and Pareto front analysis for the prediction of high-performance candidate materials for electrocatalysis in regions where other data sources are scarce or non-existent.
Candidate compositions are evaluated on the basis of their similarity to the terms `conductivity' and `dielectric', which enables reaction-specific candidate composition predictions for oxygen reduction (ORR), hydrogen evolution (HER), and oxygen evolution (OER) reactions.
This, combined with Pareto optimization, allows us to significantly reduce the pool of candidate compositions to high-performing compositions.
Our predictions, which are purely based on text data, match the measured electrochemical activity very well.
\end{abstract}


\section{Introduction}

Electrocatalysts are essential components in energy conversion technologies, such as fuel cells~\cite{Borup20073904,Debe201243,Lefèvre200971,Jones20241144}, water electrolyzers~\cite{Zaman20242922,Mandal20228611,Fujigaya201710584}, and metal-air batteries~\cite{Milikić2023,Salado2022,Ye2020364}.
These technologies are critical for transitioning to a clean energy future, as they enable efficient energy storage and conversion without reliance on fossil fuels.
At the heart of these technologies are reactions such as the oxygen reduction reaction (ORR)~\cite{Statt20231078,Gong2009760,Okubo20242183}, hydrogen evolution reaction (HER)~\cite{Talledo20241430,Greeley2006909}, and oxygen evolution reaction (OER)~\cite{Abed20242265,McCrory201316977}.
Each of these reactions plays an important role in processes like electricity generation and hydrogen production, but their efficiency depends on the performance of the electrocatalysts involved.
Developing high-performance, cost-effective electrocatalysts is therefore the basis for viable use cases.

However, the discovery and design of new electrocatalysts presents significant challenges.
Electrocatalysts are often composed of multiple elements~\cite{Batchelor2019834,ZERDOUMI2024101590}, and their performance is influenced not only by the presence of specific elements but also by their precise proportions and interactions~\cite{Khan2017622,Hossain202221583}.
This creates a practically infinite search space, where the number of possible compositions far exceeds what can be screened experimentally.
In addition, each reaction type poses unique requirements.
For example, ORR and HER demand materials with high electrical conductivity~\cite{Kumar2022211,Shinde2016173} to support fast charge transfer, while OER benefits from materials with higher dielectric properties to promote oxygen evolution~\cite{Wang20225350,Thao2023}.
Balancing these often-opposing properties further complicates the search for optimal reaction-specific materials.

Traditional methods for the discovery of electrocatalysts are based on trial and error experiments~\cite{ZHANG2022140553,She2017}.
Researchers typically test individual compositions one at a time, measuring their performance under specific conditions.
Although this approach has led to many important breakthroughs, it is slow, expensive, and inefficient, particularly as the (compositional) complexity of material systems grows.
Even high-throughput approaches alone can not tame the `combinatorial explosion' to tune and optimize the precise compositions and make use of the available design space~\cite{Loeffler2021a}.
Therefore, many potentially promising materials remain untested, either because they are overlooked or because testing every possibility is simply impractical.
This represents a bottleneck in the discovery and optimization of future energy technologies.

Experimental data is precious because it is expensive to obtain even in high-throughput, automated labs.
Quantum-accurate simulations also require many calculations in quaternary or quinary composition spaces when high resolution in composition space is required~\cite{Stricker2025a}.
Knowledge in scientific texts, however, exists.
The challenge is how to use it effectively.
Thousands of scientific articles are published every year that contain potentially valuable data and insights on material properties, synthesis methods, and performance metrics. 
However, this information is scattered and difficult to integrate into a cohesive \textit{picture} to establish composition-property relationships.
The lack of systematic methods to extract and leverage this knowledge has probably left many materials underexplored or just overlooked.
Bridging this gap requires tools that can process and analyze large volumes of scientific literature and models that use this information to establish correlative structure-property relationships.

Advances in data science, particularly natural language processing (NLP)~\cite{Lei20241257,Muthukkumaran2023} and machine learning~\cite{Liu2022,Chen2022,Mai202213478}, offer promising solutions to these challenges.
NLP techniques can process large amounts of text, e.g. a corpus of research articles, extracting correlations between material compositions, properties, and performance metrics, e.g. in the form of word embeddings.
Word embeddings, like those generated by Word2Vec models, encode complex materials-specific terms as numerical representations amenable for model development and optimization.
The relationship between composition and property vectors enables quantitative comparisons that can be used predictively~\cite{Zhang2024}.
These techniques not only automate the extraction of latent knowledge from text but also provide a way to link material representations to measurable properties~\cite{Tshitoyan2019} like conductivity or dielectric behavior (polarizability), purely based on the probability of co-occurrence of words as they are presented in scientific texts.
Such links enable the evaluation of potential applications of a material prior to any experimental measurement, thereby opening the door to systematically accelerate material discovery.

Our study introduces an approach that combines automated text mining, Word2Vec models, and Pareto optimization to filter likely high-performing candidate compositions from a large number of possible compositions for application in the three electrocatalytic reactions mentioned above.
Text mining extracts relevant information from open-access abstracts; Word2Vec models are used to generate embeddings for the calculation of similarity scores for key properties.
These scores are then used in Pareto optimization to filter large datasets for best-performing candidate materials.
Our approach reduces experimental workload by leveraging reaction-specific Pareto optimization directions tailored to ORR, HER, and OER.
By linking qualitative descriptions from scientific literature to quantitative optimization, our method offers a scalable framework for accelerated materials discovery and optimization.

\section{Methods}

We describe in detail our workflow in the following: From the collection of open-access abstracts of scientific articles, Word2Vec model building, Pareto front calculation.
All code to reproduce our findings or reuse our approach is publicly accessible and licensed under LGPL in Version 3.0~\cite{Zhang2025}.

\subsection{Automated Paper Collection}

To create a Word2Vec embeddings~\cite{Mikolov2013,Goldberg2014}, abstracts related to electrocatalysts and high-entropy alloys are collected using the \texttt{PaperCollector} module in \texttt{MatNexus}~\cite{Zhang2024}.
The sources of these abstracts are Scopus and ArXiv.
In our study we only include only open-access publications.
To keep the data set comprehensive and up-to-date, the query includes all relevant publications up to the year 2024.
The query includes automated retrieval of bibliographical metadata along with the respective abstracts.
The collected data is then saved in a structured CSV format before further processing.

\subsection{Text Processing}

Text processing uses the \texttt{TextProcessor} module of \texttt{MatNexus}.
This step preprocesses the collected raw abstracts for natural language processing by cleaning and organizing the text.

\textit{Cleaning} and filtering comprises:
\begin{itemize}
    \item \textbf{Removing unnecessary content:} Parts of the text containing “©” or “\& Co.” are removed to exclude licensing information and publisher-specific content.
    \item \textbf{Filtering out common words:} Standard English stopwords are removed to reduce noise, ensuring that only meaningful words are kept.
    \item \textbf{Extracting chemical information:} Chemical element symbols and formulas are identified and kept, as they are key to establishing composition-property relationships in word embedding space.
\end{itemize}

The cleaned text is tokenized and saved in a structured format, ready for model training.

\subsection{Word2Vec Model Training}

To train a Word2Vec model~\cite{Mikolov2013,Goldberg2014}, we use the \texttt{VecGenerator} module in \texttt{MatNexus}.
Compared to other word vector training methods, such as count-based embeddings or TF-IDF, Word2Vec generates dense vectors that represent words in a continuous vector space, allowing for an assessment of the correlations of terms which is important for our method.
Unlike methods that rely solely on word co-occurrence frequencies, Word2Vec considers the context within sentences: words appearing in similar contexts are mapped to close-by vector representations measured by cosine similarity.
More advanced techniques like transformers (e.g., BERT or GPT) for understanding text exist; however, they require significantly larger computational resources and training datasets to perform effectively, which can be a limiting factor when working with domain-specific corpora like ours.
Additionally, transformers are often overparameterized for tasks such as word embedding generation, where simpler models such as Word2Vec are sufficient to capture the essential relationships~\cite{Rogers2020842}.
Given the specific goal of creating word embeddings to support material discovery, we chose Word2Vec for its balance between computational efficiency and accuracy for our scenario.

For training, we selected the following parameters:
\begin{itemize}
    \item \textbf{Model type:} Skip-gram was used to capture contextual relationships in texts.
    \item \textbf{Vector size:} The word vectors were set to 200 dimensions to represent the information.
    \item \textbf{Training method:} Hierarchical softmax was used to handle large vocabulary.
    \item \textbf{Context window:} A window size of 5 words was used to capture local relationships between terms.
    \item \textbf{Minimum word frequency:} Words appearing at least once were included to maximize the dataset's coverage.
    \item \textbf{Parallel processing:} Multiple threads were used to for training efficiency.
\end{itemize}

The processed abstracts are used to train the Word2Vec model with these settings to create word embedding-based representations for downstream analysis.

\subsection{Dataset Preparation and Similarity Calculation}

The dataset preparation process involves loading a pre-trained Word2Vec model as per the previous explanation.
With this we calculate material similarities for target properties such as `dielectric' and `conductivity'.
These properties were chosen for their relevance to the study of electrocatalysts and high-entropy alloys~\cite{Foppa20211016}.

The input data consisted of many different compositions for different material systems.
`Material system' refers to a fixed number elements.
Within one material system, only the concentration (composition) varies.
For each specific composition, similarity scores w.r.t. the target properties are calculated using the Word2Vec model.
The resulting similarity scores indicate how \textit{similar} the embedding representation of a specific compositions is with a selected property.
Similarity scores are then used as features for each composition for the Pareto front optimization as per the following step.

\subsection{Pareto Front Calculation}

The Pareto front method is a widely used multi-objective optimization technique that identifies solutions offering the best trade-offs between conflicting objectives~\cite{pareto1896cours}.
This method has been applied in various fields, such as materials design and engineering~\cite{Djeffal201329,Bale20222100}, where multiple competing goals are optimized.
In Pareto optimization, a solution is considered non-dominated if no other solution performs better in all objectives simultaneously.
By focusing on non-dominated solutions, Pareto optimization narrows the search space to candidates that strike a balance between the objectives, enabling decision-making in complex systems.
We use Pareto front optimization to identify optimal trade-offs between material properties tailored for specific reactions: hydrogen evolution reaction (HER), oxygen reduction reaction (ORR), and oxygen evolution reaction (OER).
The objectives and optimization directions were defined based on the reaction type, reflecting the opposing nature of these processes.

\begin{figure}
    \centering
    \includegraphics[width=0.6\textwidth]{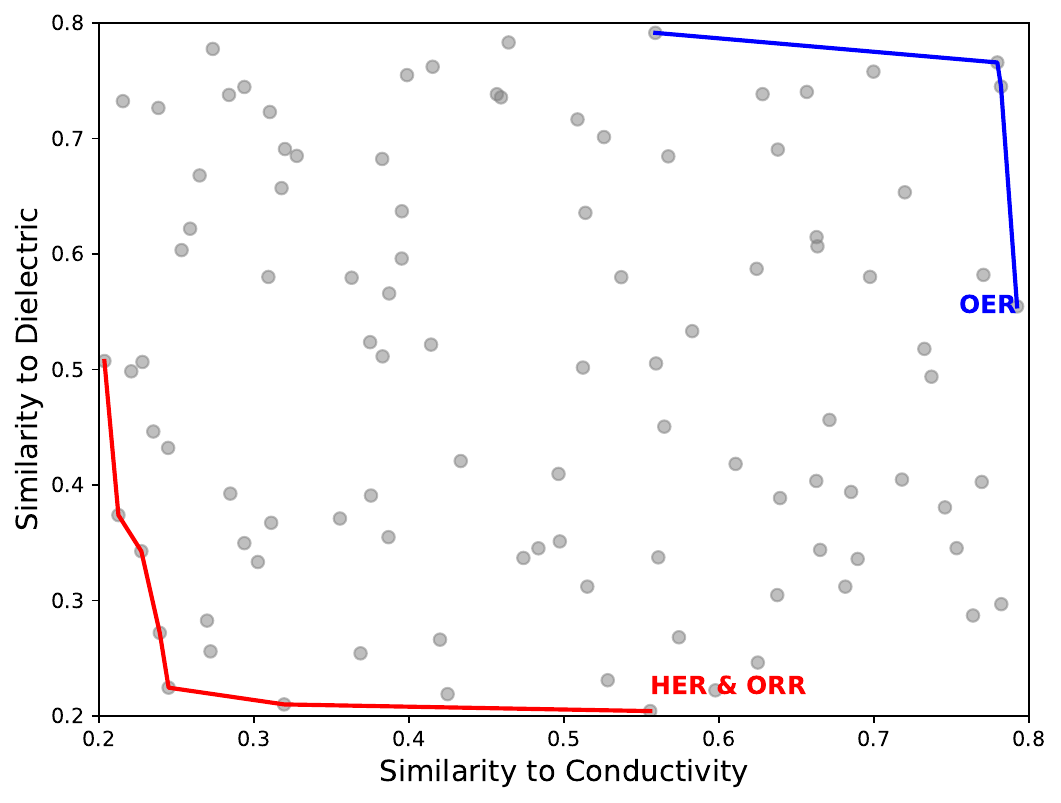}
    \caption{Pareto front visualization depicting trade-offs between similarity to conductivity and dielectric properties.}
    \label{fig:pareto_illustration}
\end{figure}

Fig.\ref{fig:pareto_illustration} schematically illustrates the Pareto front for HER/ORR and OER w.r.t the similarity of compositions with the terms `conductivity' and `dielectric'.
Gray points represent randomly generated data points for the two similarities.
For HER and ORR, the goal is to maximize similarity to `conductivity' and minimize similarity to `dielectric' (red line), as high conductivity and low dielectric properties are desirable~\cite{Kumar2022211,Shinde2016173}.
In contrast, for OER, the goal is to minimize similarity to `conductivity' and maximize similarity to `dielectric' (blue line)~\cite{Wang20225350,Thao2023}.
For each material system, we create such a dielectric-conductivity-similarity space.
Within each of these spaces, the compositions on the Pareto front are our prediction of promising compositions suggested for experimental validation.

\section{Results}

\begin{table*}
\small
\caption{Comparative elemental composition in \% across systems: all compositions considered and Pareto-optimal compositions.}
\label{tbl:elementalConcentrations}
\begin{tabular*}{\textwidth}{@{\extracolsep{\fill}}lccccccc}
\hline
System & Element & Max (Original) & Max (Pareto) & Min (Original) & Min (Pareto) & Std Dev (Original) & Std Dev (Pareto) \\
\hline
AgPdPt & Pt & 68 & 58 & 17 & 41 & 13.136 & 6.01 \\
           & Pd & 46 & 44 & 0  & 40 & 12.732 & 1.57 \\
           & Ag & 69 & 14 & 1  & 1  & 17.097 & 4.439 \\
\hline
AgPdRu & Pd & 86 & 86 & 22 & 65 & 16.448 & 7.934 \\
           & Ag & 40 & 14 & 9  & 9  & 7.808  & 1.242 \\
           & Ru & 45 & 22 & 0  & 0  & 12.393 & 8.484 \\
\hline
AgPdPtRu & Pt & 56 & 55 & 0  & 50 & 14.901 & 2.928 \\
               & Pd & 27 & 24 & 0  & 21 & 7.325  & 1.753 \\
               & Ag & 39 & 12 & 3  & 7  & 9.299  & 2.422 \\
               & Ru & 67 & 17 & 7  & 10 & 15.388 & 3.589 \\
\hline
AgAuPdPtRh & Pt & 23 & 21 & 2  & 5  & 5.284  & 6.889 \\
                   & Rh & 38 & 38 & 5  & 6  & 8.219  & 15.751 \\
                   & Pd & 37 & 36 & 6  & 7  & 8.463  & 13.69 \\
                   & Au & 52 & 25 & 9  & 19 & 11.447 & 2.645 \\
                   & Ag & 45 & 28 & 7  & 14 & 10.081 & 6.074 \\
\hline
AgAuPdPtRu & Pt & 42 & 34 & 7  & 24 & 9.392  & 3.017 \\
                   & Pd & 45 & 43 & 10 & 38 & 9.061  & 1.785 \\
                   & Au & 15 & 9  & 2  & 6  & 3.259  & 1.154 \\
                   & Ag & 25 & 7  & 3  & 5  & 5.309  & 0.741 \\
                   & Ru & 61 & 16 & 12 & 12 & 13.384 & 1.518 \\
\hline
NiPdPtRu & Pt & 90 & 13 & 0  & 0  & 25.543 & 2.426 \\
               & Pd & 89 & 10 & 0  & 0  & 25.749 & 1.771 \\
               & Ni & 80 & 80 & 0  & 5  & 18.511  & 17.727 \\
               & Ru & 94 & 94 & 0  & 8  & 26.098 & 20.515 \\
\hline
\end{tabular*}
\end{table*}

\begin{table*}[h!]
\small
\caption{Measured current densities (mA/cm\textsuperscript{2}) and statistical details for electrocatalysts under specified potentials}
\label{tbl:currentDensityMetrics}
\begin{tabular*}{\textwidth}{@{\extracolsep{\fill}}lccccccccc}
\hline
\textbf{System} & \textbf{Potential (mV)} & 
\begin{tabular}[c]{@{}c@{}} \textbf{Max} \\ \textbf{(Original)} \end{tabular} & 
\begin{tabular}[c]{@{}c@{}} \textbf{Max} \\ \textbf{(Pareto)} \end{tabular} & 
\begin{tabular}[c]{@{}c@{}} \textbf{Min} \\ \textbf{(Original)} \end{tabular} & 
\begin{tabular}[c]{@{}c@{}} \textbf{Min} \\ \textbf{(Pareto)} \end{tabular} & 
\begin{tabular}[c]{@{}c@{}} \textbf{Mean} \\ \textbf{(Original)} \end{tabular} & 
\begin{tabular}[c]{@{}c@{}} \textbf{Mean} \\ \textbf{(Pareto)} \end{tabular} & 
\begin{tabular}[c]{@{}c@{}} \textbf{Std Dev} \\ \textbf{(Original)} \end{tabular} & 
\begin{tabular}[c]{@{}c@{}} \textbf{Std Dev} \\ \textbf{(Pareto)} \end{tabular} \\
\hline
AgPdPt & 850 & -0.063 & -0.415 & -0.583 & -0.447 & -0.342 & -0.430 & 0.098 & 0.011 \\
AgPdRu & 850 & -0.065 & -0.364 & -0.673 & -0.673 & -0.278 & -0.463 & 0.114 & 0.086 \\
AgPdPtRu & 850 & -0.060 & -0.361 & -0.366 & -0.366 & -0.159 & -0.362 & 0.074 & 0.003 \\
AgAuPdPtRh & -300 & -0.688 & -0.982 & -1.128 & -1.085 & -0.892 & -1.037 & 0.098 & 0.043 \\
AgAuPdPtRu & -300 & -0.794 & -1.343 & -1.494 & -1.437 & -1.333 & -1.395 & 0.107 & 0.030 \\
NiPdPtRu & 1700 & 6.896 & 6.896 & 0.237 & 0.756 & 1.073 & 3.265 & 0.843 & 1.633 \\
\hline
\end{tabular*}
\end{table*}

\subsection{Overview of the Datasets}

We investigate six different material systems from the class of compositionally complex solid solutions (a.k.a. high-entropy alloys)~\cite{ZERDOUMI2024101590}.
These systems are classified  in three groups for the three reaction types: the oxygen reduction reaction (ORR), hydrogen evolution reaction (HER), and oxygen evolution reaction (OER).
The ORR group contains the AgPdPt, AgPdRu, and AgPdPtRu systems, the HER group contains the two systems AgAuPdPtRh and AgAuPdPtRu, and the OER group OER group is represented by the NiPdPtRu system.
In the following, we present an overview of the experimentally-measured datasets along our predictions for each reaction type along our predictions.

\subsection{Results for ORR Systems}

\subsubsection{Elemental Composition}

Table~\ref{tbl:elementalConcentrations} provides a statistical overview of the ORR systems for both the original variance in the data as well as the variance of our Pareto-optimal predictions.
A full version of the experimental dataset is published on Zenodo~\cite{Banko2024}.
The compositions at the Pareto front exhibit a much narrower range of elemental concentrations compared to the original datasets. For instance:
\begin{itemize}
    \item In the AgPdPt system, the maximum concentration of Pt reduced from 68\,\% to 58\,\%, while the minimum increased from 17\,\% to 41\,\%. This suggests that high-performing candidates favor a balanced composition of Pt.
    \item For AgPdRu, the minimum concentration of Pd increased significantly from 22\,\% to 65\,\%, suggesting its importance for ORR activity.
\end{itemize}

\subsubsection{Current Density Metrics}

\begin{figure*}
    \centering
    \includegraphics[width=1\linewidth]{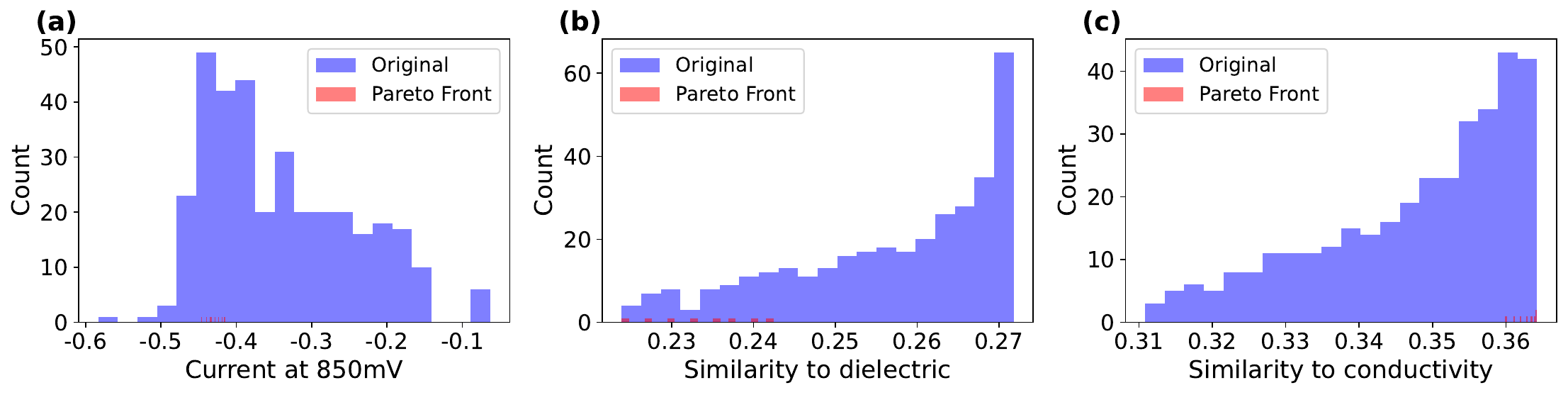}
    \caption{Measured current density distribution of AgPdPt system with their respective Pareto fronts highlighted: (a) current at 850\,mV, (b) similarity to `dielectric', and (c) similarity to `conductivity'.}
    \label{fig:Ag_Pd_Pt_distribution}
\end{figure*}

\begin{figure*}
    \centering
    \includegraphics[width=1\linewidth]{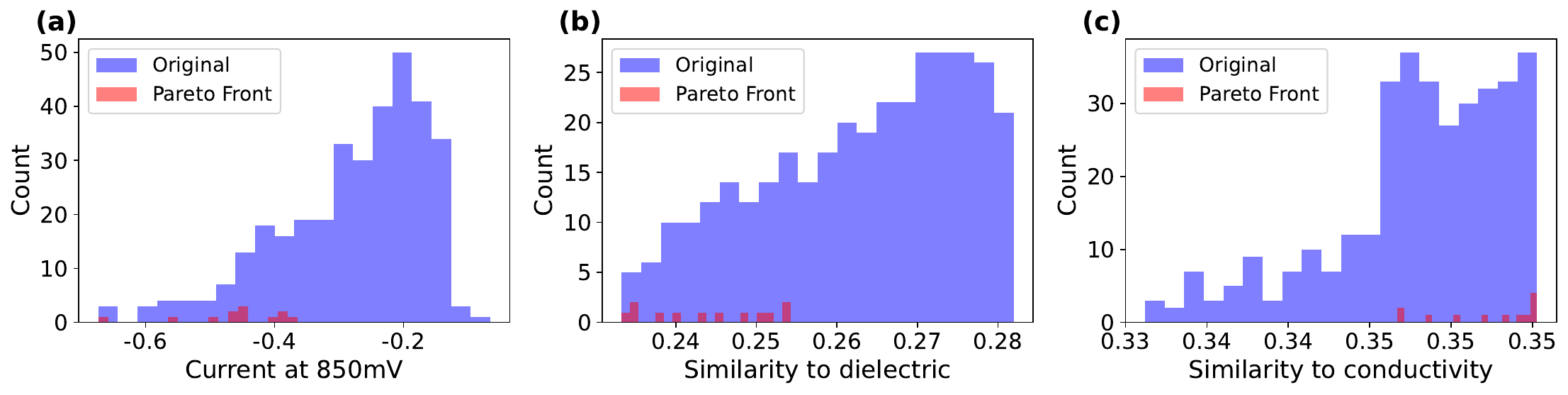}
    \caption{Measured current density distribution of AgPdRu system with their respective Pareto fronts highlighted: (a) current at 850\,mV, (b) similarity to `dielectric', and (c) similarity to `conductivity'.}
    \label{fig:Ag_Pd_Ru_distribution}
\end{figure*}

\begin{figure*}
    \centering
    \includegraphics[width=1\linewidth]{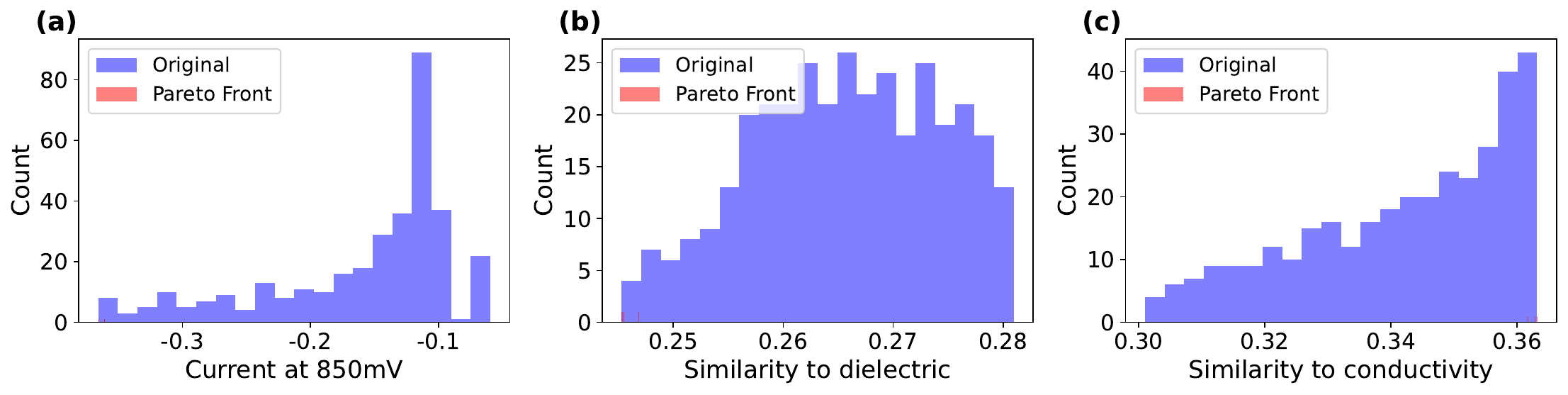}
    \caption{Measured current density distribution of AgPdPtRu system with their respective Pareto fronts highlighted: (a) current at 850\,mV, (b) similarity to `dielectric', and (c) similarity to `conductivity'.}
    \label{fig:Ag_Pd_Pt_Ru_distribution}
\end{figure*}

Fig.~\ref{fig:Ag_Pd_Pt_distribution}, \ref{fig:Ag_Pd_Ru_distribution}, and \ref{fig:Ag_Pd_Pt_Ru_distribution} present the measured current density and similarity scores distribution for the AgPdPt, AgPdRu, and AgPdPtRu systems, respectively. Each figure consists of three panels: (a) shows the distribution of current density values, while (b) and (c) display the distribution of similarity scores to `dielectric' and `conductivity', respectively.

\begin{itemize}
    \item \textbf{Panel (a)}: The original dataset (blue) exhibits a broad distribution of current densities, ranging from highly negative to less negative. The red markers represent the Pareto front solutions, which highlight compositions with optimal trade-offs. Notably, in all three systems, the Pareto front points are concentrated toward the more negative current density regions, corresponding to better ORR performance.
    
    \item \textbf{Panel (b)}: The similarity to the term `dielectric' is shown. For all three systems, the Pareto-optimal points (red) exhibit lower similarity to `dielectric', reflecting the requirement to minimize dielectric behavior for enhanced ORR activity. The Pareto front points are selected from the lower range of the similarity score distribution.

    \item \textbf{Panel (c)}: The similarity to `conductivity' is shown. The Pareto front solutions show a higher similarity to `conductivity', aligning with the need for materials with excellent conductive properties to support fast charge transfer during ORR. This trend is observed across all three systems, where the Pareto-optimal points lie toward the higher end of the similarity score distribution.

\end{itemize}

\begin{figure*}
    \centering
    \includegraphics[width=1\linewidth]{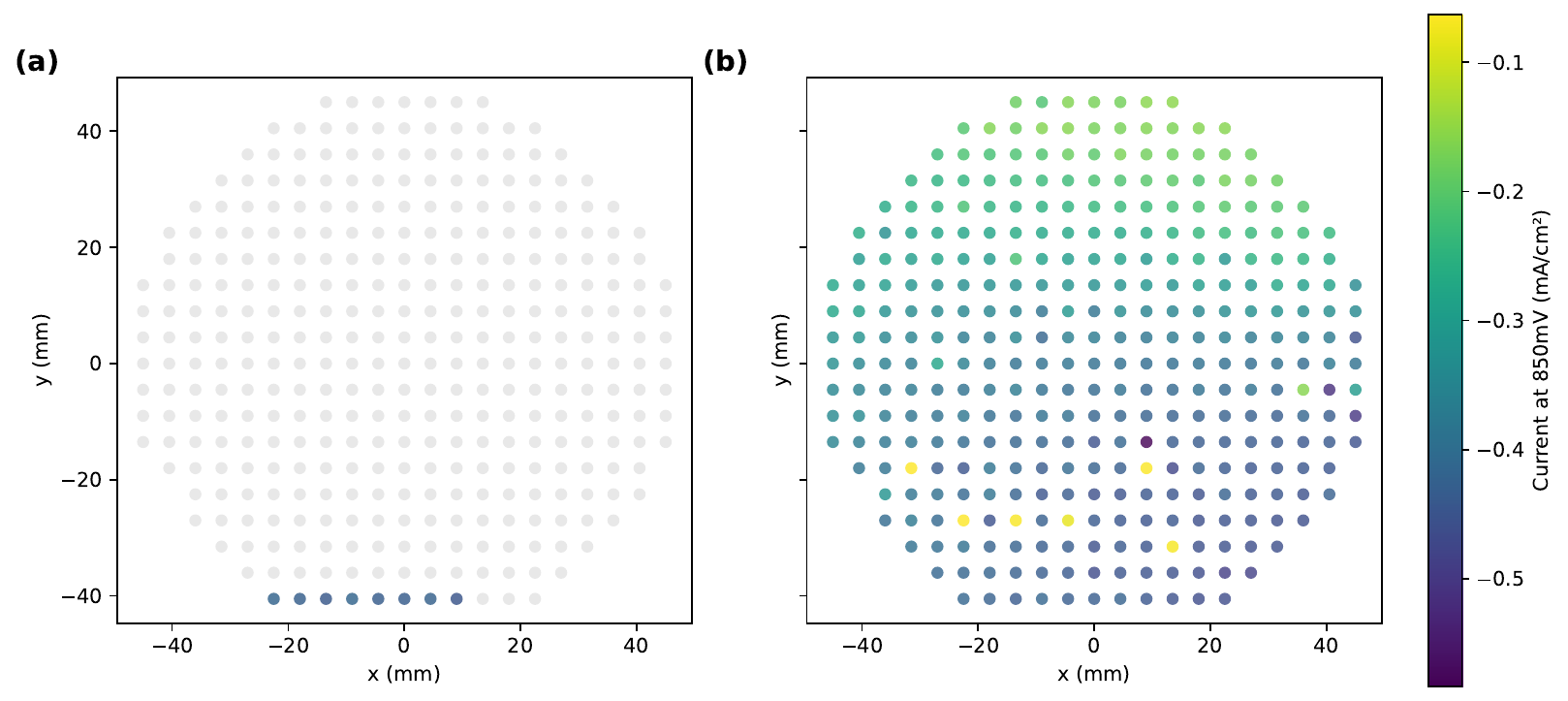}
    \caption{Measured current density for AgPdPt system: (a) High-performance predictions from Pareto front optimization highlighted and (b) original dataset with current density overlaid as color, more negative (blue) means better. Yellow locations are faulty measurements.}
    \label{fig:Ag_Pd_Pt_pareto_front}
\end{figure*}

\begin{figure*}
    \centering
    \includegraphics[width=1\linewidth]{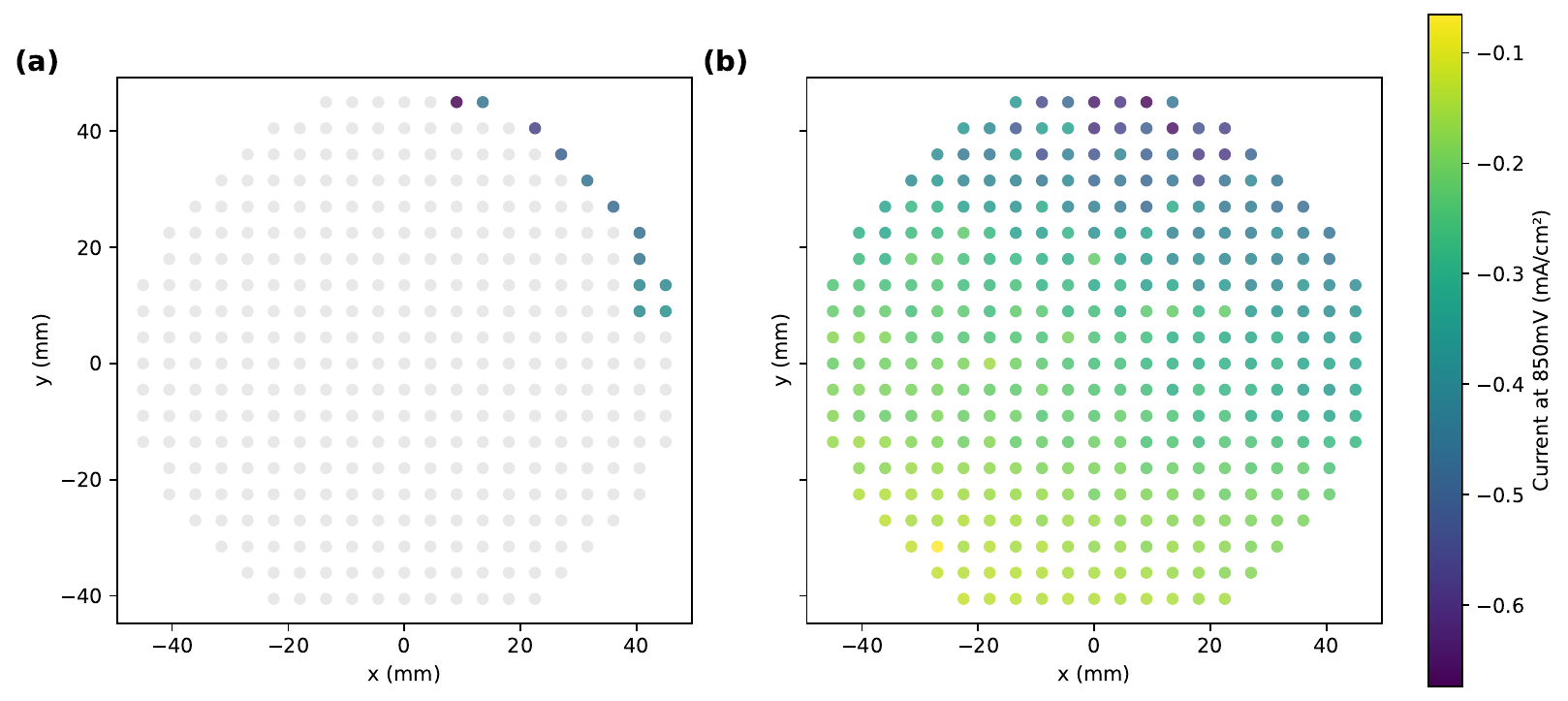}
    \caption{Like Fig.~\ref{fig:Ag_Pd_Pt_pareto_front} but for the AgPdRu system.}
    \label{fig:Ag_Pd_Ru_pareto_front}
\end{figure*}

\begin{figure*}
    \centering
    \includegraphics[width=1\linewidth]{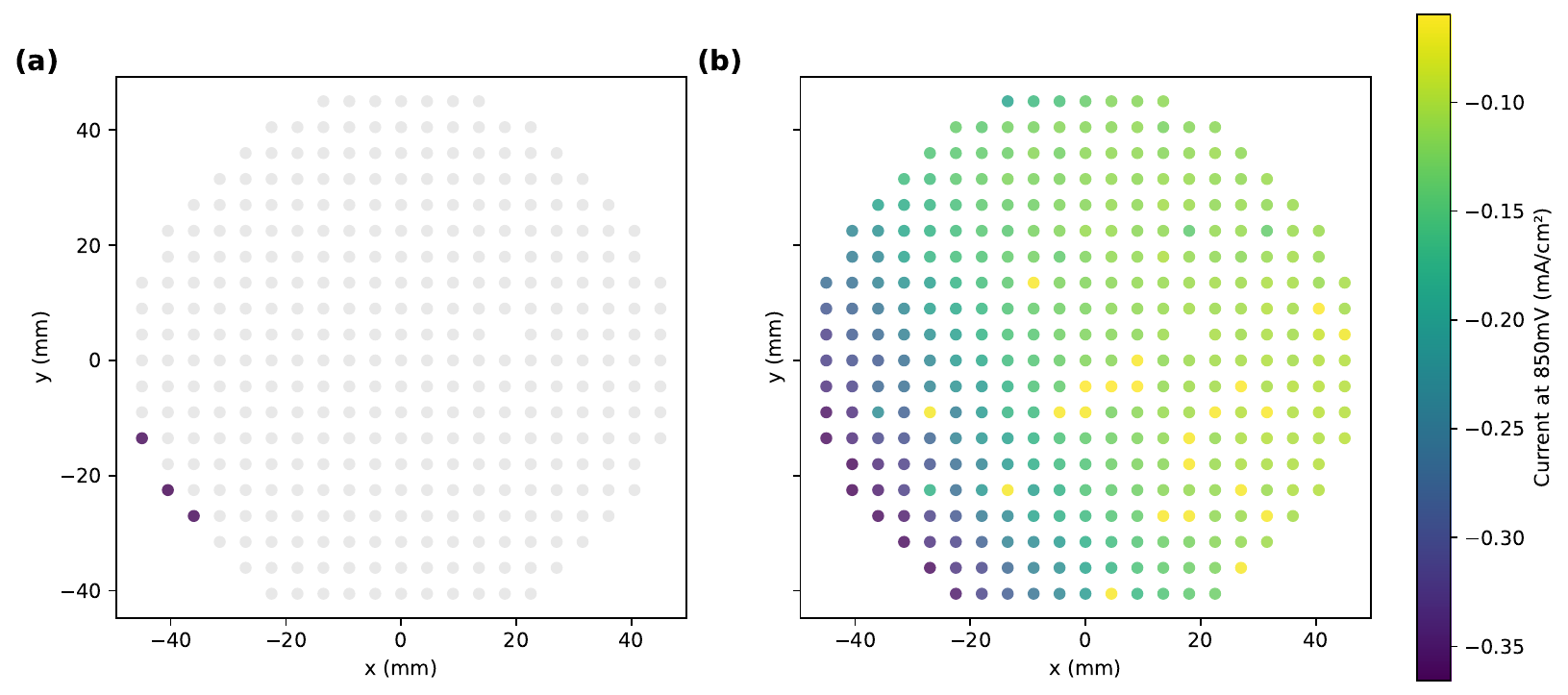}
    \caption{Like Fig.~\ref{fig:Ag_Pd_Pt_pareto_front} but for AgPdPtRu system.}
    \label{fig:Ag_Pd_Pt_Ru_pareto_front}
\end{figure*}

Fig.~\ref{fig:Ag_Pd_Pt_pareto_front}, \ref{fig:Ag_Pd_Ru_pareto_front}, and \ref{fig:Ag_Pd_Pt_Ru_pareto_front} display the measured current density for the AgPdPt, AgPdRu, and AgPdPtRu systems, respectively, at 850\,mV. 
X and y axes of the figures refer to real space coordinates of the location of measurements on a materials library -- a thin film deposited on a wafer with approximately 10\,cm diameter~\cite{Ludwig2019}.
In each figure:

\begin{itemize}
    \item \textbf{Panel (a)}: This panel shows the reduced dataset after Pareto front calculation. The remaining points represent non-dominated solutions that balance the optimization objectives, highlighting the most promising compositions for ORR. By narrowing down the dataset, the Pareto front approach reduces the search space, focusing attention on the most relevant candidates for further investigation.

    \item \textbf{Panel (b)}: Each point represents a specific material composition measured at an $(x, y)$ coordinate on a wafer. The x-axis and y-axis correspond to physical locations on the wafer, with current density values at these positions color-coded according to the scale on the right. The color gradient ranges from less negative values (yellow and green) to more negative values (blue and purple), where more negative values indicate better ORR performance. Spatial variations in current density across the wafer reflect differences in material composition or testing conditions. Some isolated points with less negative current density appear within high-performance regions, potentially due to minor measurement errors. Hollow markers indicate positions excluded due to obvious test errors, such as near-zero current density, ensuring data reliability.
\end{itemize}

The results demonstrate that the Pareto front prediction effectively identifies high-performaing compositions.
This conclusion is validated by comparing the reduced dataset in Panel (a) to the original dataset in Panel (b), where high-performing candidates selected through the Pareto front approach correspond to regions of more negative current density. 

Table \ref{tbl:currentDensityMetrics} also compares the current density performance of the original and Pareto datasets. Key observations include:
\begin{itemize}
    \item \textbf{AgPdPt}: The maximum current density improved from -0.063 mA/cm\textsuperscript{2} (original) to -0.415 mA/cm\textsuperscript{2} (Pareto), with a substantial reduction in standard deviation (0.098 to 0.011 mA/cm\textsuperscript{2}). This indicates a consistent high performance among Pareto-selected materials.
    \item \textbf{AgPdRu}: The Pareto dataset achieved a maximum current density of -0.364 mA/cm\textsuperscript{2} compared to -0.065 mA/cm\textsuperscript{2} in the original dataset, while retaining the same minimum value of -0.673 mA/cm\textsuperscript{2}. The mean current density shifted closer to higher-performing candidates.
    \item \textbf{AgPdPtRu}: The maximum current density increased from -0.060 mA/cm\textsuperscript{2} to -0.361 mA/cm\textsuperscript{2}, with the standard deviation reduced to nearly zero (0.003 mA/cm\textsuperscript{2}), indicating strong agreement among selected candidates.
\end{itemize}


\subsection{Results for HER Systems}

\subsubsection{Elemental Composition}

Table \ref{tbl:elementalConcentrations} highlights the elemental distribution in the HER systems. Notable trends include:
\begin{itemize}
    \item In the AgAuPdPtRh system, the maximum concentration of Pt decreased slightly from 23\,\% to 21\,\%, while the minimum increased from 2\,\% to 5\,\%. This suggests that balanced Pt concentrations are associated with better-performing candidates.
    \item For AgAuPdPtRu, the maximum concentration of Ru decreased significantly from 61\,\% to 16\,\%, while the minimum concentration remained constant at 12\,\%. This indicates that excessively high Ru concentrations may not be favorable for HER under the given conditions.
\end{itemize}

\subsubsection{Current Density Metrics}

\begin{figure*}
    \centering
    \includegraphics[width=1\linewidth]{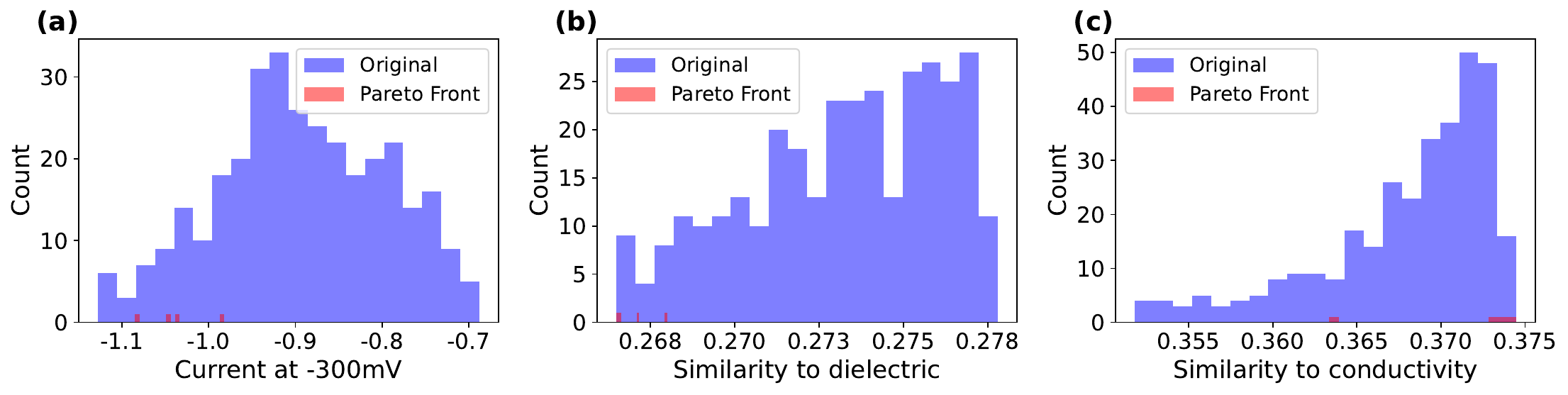}
    \caption{Measured current density distribution of AgAuPdPtRh system with their respective Pareto fronts highlighted: (a) current at -300\,mV, (b) similarity to `dielectric', and (c) similarity to `conductivity'.}
    \label{fig:Ag_Au_Pd_Pt_Rh_distribution}
\end{figure*}

\begin{figure*}
    \centering
    \includegraphics[width=1\linewidth]{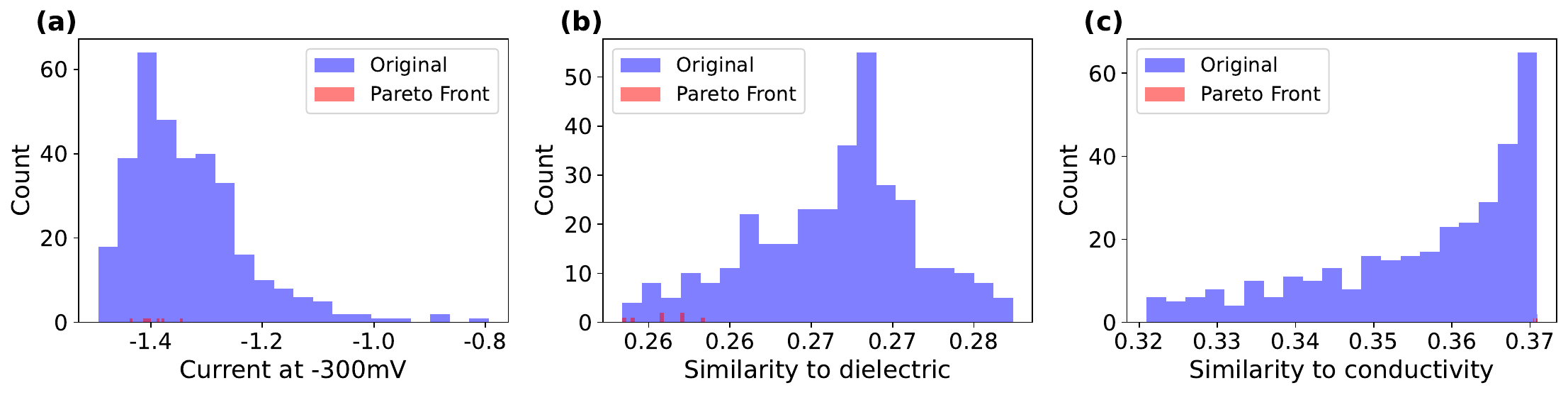}
    \caption{Measured current density distribution of AgAuPdPtRu system with their respective Pareto fronts highlighted: (a) current at -300\,mV, (b) similarity to `dielectric', and (c) similarity to `conductivity'.}
    \label{fig:Ag_Au_Pd_Pt_Ru_distribution}
\end{figure*}

Fig.~\ref{fig:Ag_Au_Pd_Pt_Rh_distribution}, \ref{fig:Ag_Au_Pd_Pt_Ru_distribution} present the measured current density and similarity scores distribution for the AgAuPdPtRh and AgAuPdPtRu systems, respectively. For these two HER material systems, the Pareto front points are concentrated toward the more negative current density regions (Panel (a)) with lower similarity score to 'dielectric' (Panel (b)) and higher similarity score to 'conductivity' (Panel (c)). 

\begin{figure*}
    \centering
    \includegraphics[width=1\linewidth]{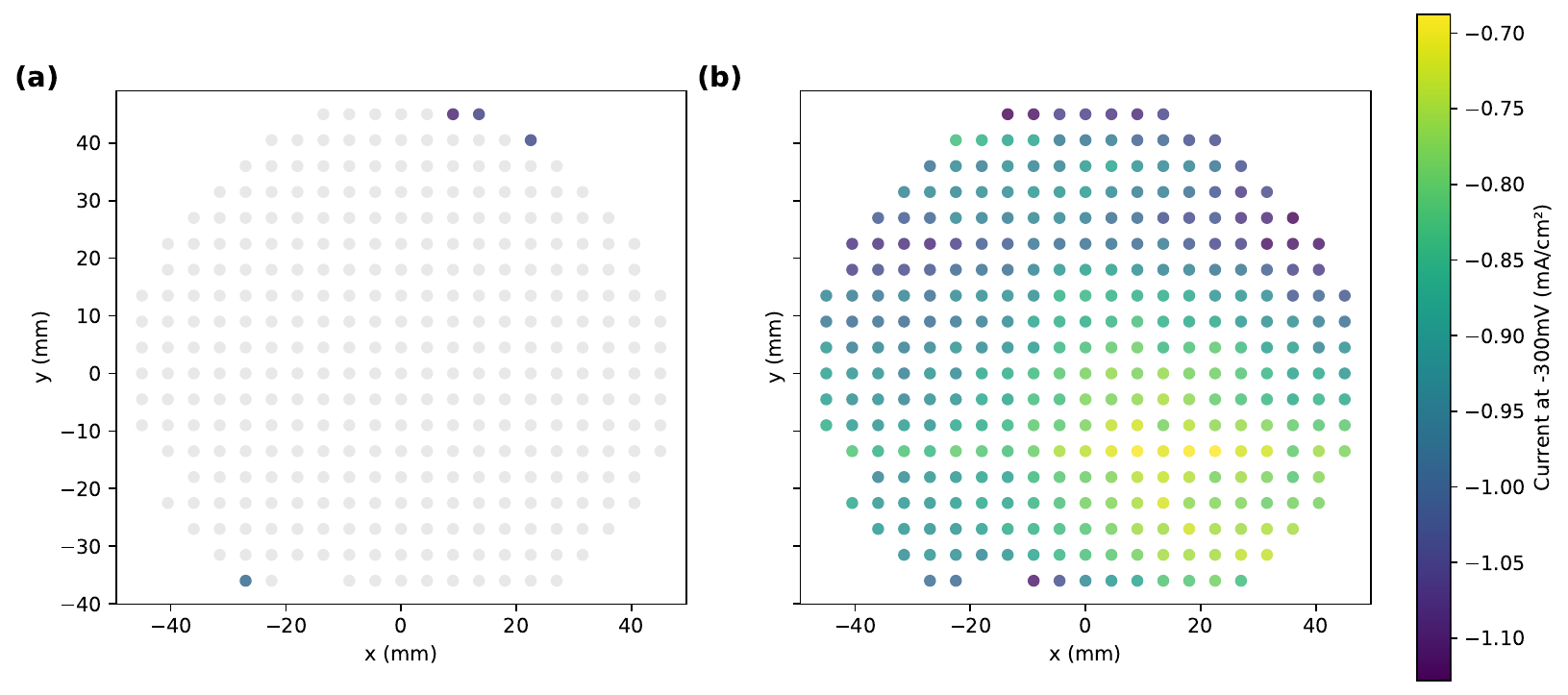}
    \caption{Like Fig.~\ref{fig:Ag_Pd_Pt_pareto_front} but for the AgAuPdPtRh system.}
    \label{fig:Ag_Au_Pd_Pt_Rh_pareto_front}
\end{figure*}

\begin{figure*}
    \centering
    \includegraphics[width=1\linewidth]{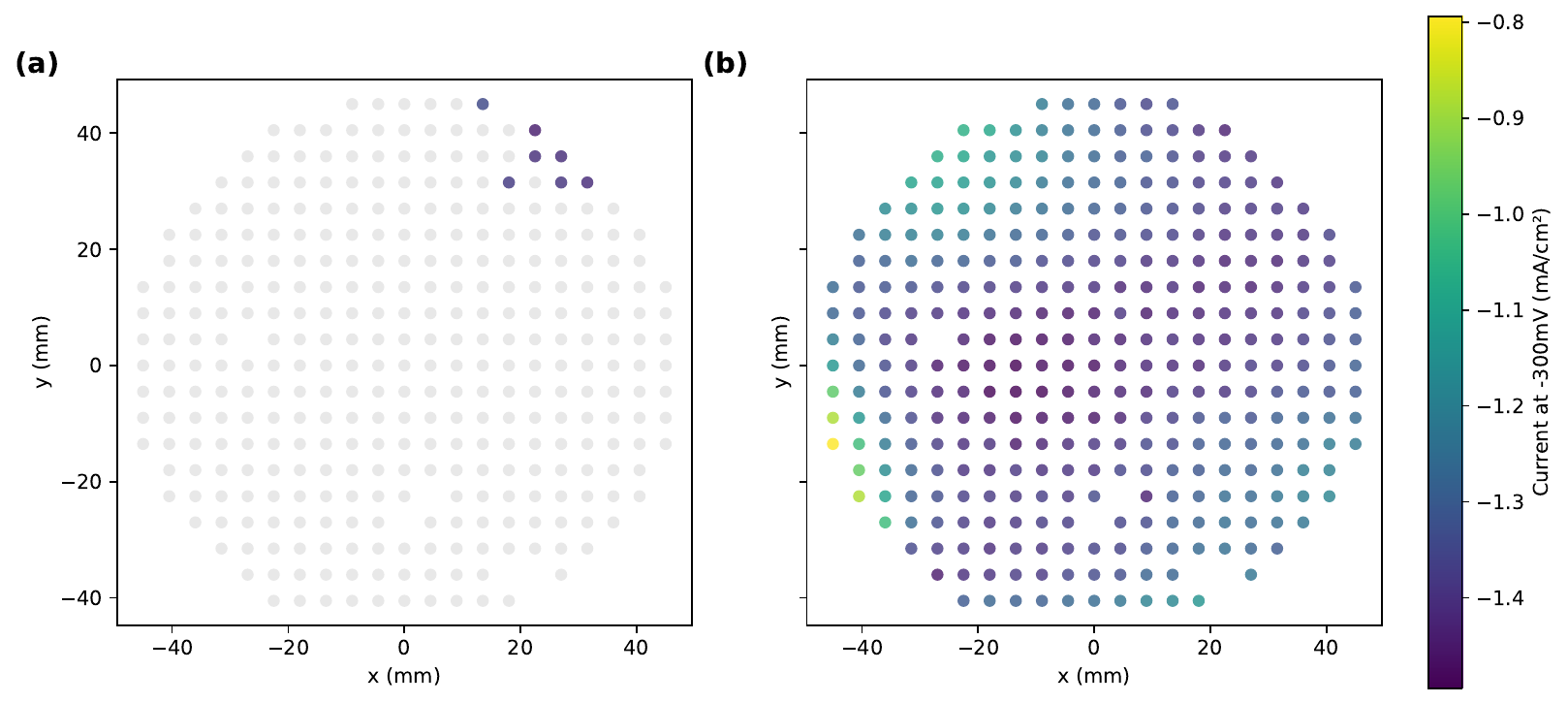}
    \caption{Like Fig.~\ref{fig:Ag_Pd_Pt_pareto_front} but for the AgAuPdPtRu system.}
    \label{fig:Ag_Au_Pd_Pt_Ru_pareto_front}
\end{figure*}

Fig.~\ref{fig:Ag_Au_Pd_Pt_Rh_pareto_front} and \ref{fig:Ag_Au_Pd_Pt_Ru_pareto_front} display the measured current density distribution after (Panel (a)) and before (Panel (b)) Pareto optimization for the AgAuPdPtRh and AgAuPdPtRu systems, respectively, at -300\,mV, relevant to the HER. 

Table~\ref{tbl:currentDensityMetrics} shows the current density performance before and after Pareto optimization. Key observations include:
\begin{itemize}
    \item \textbf{AgAuPdPtRh}: The maximum current density improved from -0.688\,mA/cm\textsuperscript{2} (original) to -0.982\,mA/cm\textsuperscript{2} (Pareto). The standard deviation decreased from 0.098 to 0.043\,mA/cm\textsuperscript{2}, indicating a consistently high performance among Pareto-selected materials.
    \item \textbf{AgAuPdPtRu}: The maximum current density improved significantly from -0.794\,mA/cm\textsuperscript{2} to -1.343\,mA/cm\textsuperscript{2}, with a reduction in standard deviation from 0.107 to 0.030\,mA/cm\textsuperscript{2}. The mean current density shifted from -1.333 to -1.395\,mA/cm\textsuperscript{2}, favoring higher-performing compositions.
\end{itemize}


\subsection{Results for OER Systems}

The experimental dataset is published on Zenodo~\cite{Thelen2025}.

\subsubsection{Elemental Composition}

Table~\ref{tbl:elementalConcentrations} shows the elemental distribution for the OER system. Notable changes after Pareto optimization include:
\begin{itemize}
    \item The maximum concentration of Pt decreased drastically from 90\% to 13\%, and its minimum concentration remained at 0\%. This indicates that lower Pt concentrations may be associated with higher-performing candidates under OER conditions.
    \item For Pd, the maximum concentration dropped from 89\% to 10\%, and its minimum concentration also remained at 0\%. This trend mirrors that of Pt, suggesting a limited influence for Pd at high concentrations in OER performance.
    \item Ni concentrations were largely unchanged, with the maximum remaining at 80\% and the minimum increasing slightly from 0\% to 5\%. This stability suggests that Ni plays a key role across a wide range of compositions.
    \item Ru concentrations showed the least variability, with the maximum remaining at 94\% and the minimum increasing slightly from 0\% to 8\%.
\end{itemize}

\subsubsection{Current Density Metrics}

\begin{figure*}
    \centering
    \includegraphics[width=1\linewidth]{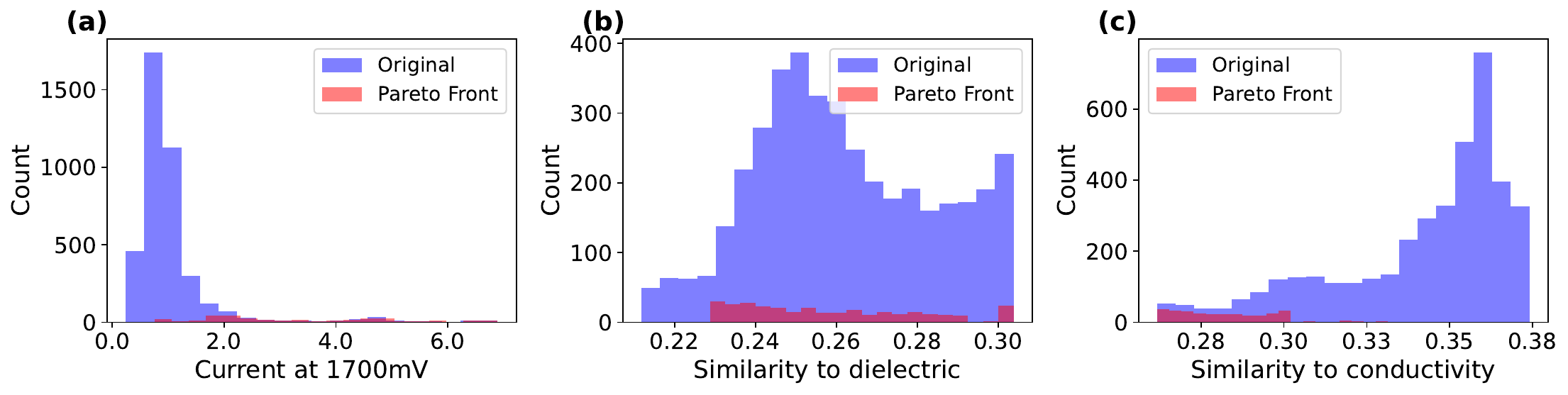}
    \caption{Measured current density distribution of NiPdPtRu system with their respective Pareto fronts highlighted: (a) current at 1700\,mV, (b) similarity to `dielectric', and (c) similarity to `conductivity'.}
    \label{fig:Ni_Pd_Pt_Ru_distribution}
\end{figure*}

Fig.~\ref{fig:Ni_Pd_Pt_Ru_distribution} shows the measured current density at 1700\,mV and similarity scores distribution for the NiPdPtRu material system.
The compositions selected through Pareto optimization are almost all distributed in higher current density region, which indicates better OER performance (Panel (a)), with high similarity score to 'dielectric' (Panel (b)) and low similarity score to 'conductivity' (Panel (c)). 

\begin{figure*}
    \centering
    \includegraphics[width=\textwidth]{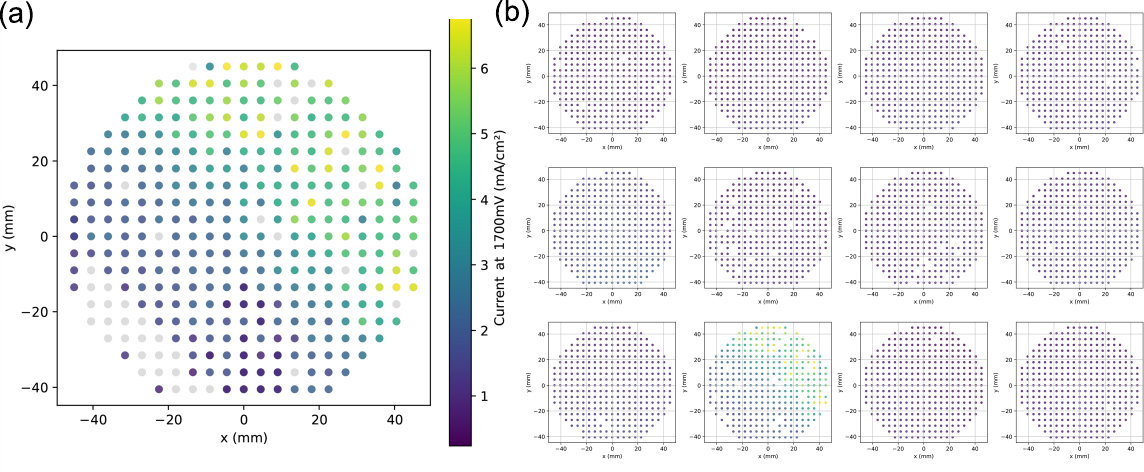}
    \caption{Measured current density for the NiPdPtRu system: (a) High-performance predictions from Pareto front optimization highlighted and (b) original dataset spread across 12 experimental samples with current density overlaid as color, more positive (yellow) means better.}
    \label{fig:Ni_Pd_Pt_Ru_pareto_front}
\end{figure*}

Figure~\ref{fig:Ni_Pd_Pt_Ru_pareto_front} shows the measured current density for the NiPdPtRu system at 1700\,mV.
In this system, data was collected from 12 individual experimental samples which span the composition range.
Panel (a) displays the results after Pareto front calculation and Panel (b) shows the full data set across 12 individual samples.
Individual subplots correspond to different wafers, illustrating the variations in current density due to compositional differences.
Our method successfully identifies all the best-performing compositions which were measured on the sample displayed in the top right.
This result demonstrates the method's capability to effectively isolate high-performance candidates from a large dataset with spatial and compositional variations.
All but the one identified sample show lower current density values.

As shown in Table~\ref{tbl:currentDensityMetrics} and Fig.~\ref{fig:Ni_Pd_Pt_Ru_distribution}, the performance of the NiPdPtRu system improved significantly after Pareto optimization:
\begin{itemize}
    \item The mean current density increased from 1.073\,mA/cm\textsuperscript{2} (original) to 3.265\,mA/cm\textsuperscript{2} (Pareto), reflecting a shift toward higher-performing candidates.
    \item The minimum current density improved from 0.237\,mA/cm\textsuperscript{2} to 0.756\,mA/cm\textsuperscript{2}, further demonstrating the effectiveness of Pareto optimization.
    \item The standard deviation increased from 0.843\,mA/cm\textsuperscript{2} (original) to 1.633\,mA/cm\textsuperscript{2} (Pareto), indicating greater variability among the selected candidates, potentially due to the broader concentrations of elements in the original dataset.
\end{itemize}

\section{Discussion}

Our approach, which combines text mining, word embedding-based representations of compositions and properties in conjunction with Pareto front optimization, demonstrates its effectiveness by systematically narrowing down high-performing composition candidates from large compositional spaces with a very fine resolution in individual element concentrations at the level of single percentage changes.
We automate the extraction of materials-relevant information from scientific literature, transforming qualitative descriptions into quantitative metrics.
The use of word embeddings to represent materials and their properties, such as conductivity and dielectric behavior, allows us to evaluate candidate compositions w.r.t. their experimentally-measured performance.
Our method provides a tool to significantly reduce the experimental workload by focusing on compositions which are likely to meet specific performance criteria.
Compared to trial-and-error approaches and even high-throughput screening, our strategy accelerates the identification of promising materials for electrocatalytic applications, offering a scalable framework that bridges data science and materials discovery and optimization and thereby directly addresses the problem of the `combinatorial explosion'.

However, there are several assumptions in our approach which warrant further discussion.
First, we assume that the multi-element compositions can be represented as a weighted linear combination of single-element representations.
The main reason to linearly combine single-element representations is that almost none of the specific compositions for compositionally complex solid solutions have been measured experimentally and subsequently reported.
This is in contrast and the main difference to the one-hot encoding of materials with several elements, e.g.~\cite{Tshitoyan2019} where materials like $\textrm{Bi}_2\textrm{Te}_3$ receive their own embedding.
In our approach, individual elements -- Bi and Te in this example -- are individually encoded.
Thereby, we open up the representation space for arbitrary compositions which have not been explicitly reported in literature.
While this approximation is computationally straightforward to represent arbitrary compositions, it does not account for potential non-linear effects or complex interactions between elements.
Given the success of our approach, we assume that possible non-linear relationships are to some extent taken care of by the high dimensionality (200) of our embedding space.
Linearly combining individual elements' representations for arbitrary composition consequently also implies that composition-property relationships are smooth in the high-dimensional embedding space.
We argue that this is a valid assumption since many material properties vary comparatively smoothly with composition.

Second, the use of similarity scores for `conductivity' and `dielectric' as optimization objectives is based on the known requirements for specific electrochemical reactions (e.g., high conductivity for ORR and HER, high dielectric for OER).
This assumption is based on domain knowledge and does not generalize to other applications without modification.
In other words, the similarity scores with specific materials properties as descriptors for reaction-specific performance predictions can be seen as domain knowledge informed \textit{lucky guesswork}.
For other reactions or material classes or even completely other problems, alternative target properties (`descriptors') must be identified based on experimental insights or theoretical considerations.

Lastly, Pareto front optimization is to identify high-performance candidate compositions. While the current implementation demonstrates the flexibility of Pareto optimization, additional objectives such as stability, cost, or scalability could be included to refine the predictions further.
Different multi-objective optimization strategies could also be explored to better tailor the results for specific reactions or material systems.

By addressing these assumptions and exploring potential refinements, the proposed framework can be extended to a broader range of materials discovery and design challenges to include more constraints.
The results presented here highlight the power of integrating latent knowledge in scientific texts with an optimization algorithm, showcasing a path toward data-driven, data-efficient exploration of complex compositional spaces where little prior data, specifically application-relevant data, exists.

\subsection{Comparison Across Reactions}

The analysis of ORR, HER, and OER systems highlights the effectiveness of combining text mining, word embeddings, and Pareto optimization to identify high-performance electrocatalysts reliably without any prior knowledge about their electrocatalytic performance.
Each reaction requires a different strategy.
This section discusses the broader implications of the observed trends.

\begin{itemize}
    \item \textbf{ORR systems}: Narrower ranges of Pt and Pd concentrations were observed, indicating a need for balanced compositions to optimize catalytic activity. This reflects the importance of synergy between these elements for efficient oxygen reduction.
    \item \textbf{HER systems}: Moderate concentrations of Ru and Pt were favored, while excessively high concentrations were filtered out. This suggests that a balanced distribution of these key elements is more effective for HER than a high concentration of a single element in large amounts, likely because balanced compositions provide a better environment for catalytic activity across the composition space.
    \item \textbf{OER systems}: The broader range of Ni and Ru concentrations retained after optimization suggests that OER can tolerate diverse compositions. This may indicate that the performance of OER systems is less dependent on precise elemental ratios and more on the overall presence of these active elements.
\end{itemize}

These trends demonstrate that elemental composition preferences are reaction-specific, emphasizing the importance of tailoring optimization objectives to the unique electrochemical demands of each reaction~\cite{Kumar2022211,Shinde2016173,Wang20225350,Thao2023}.

\subsection{Role of Text Mining and Word Embeddings in Optimization}

Text mining and word embeddings provided a systematic way to incorporate material representations into an optimization or discovery process.
By processing scientific abstracts, text mining extracted relevant terms and relationships that are often scattered across literature.
Word embeddings transform these terms into numerical representations, making it possible to calculate similarity metrics for properties like conductivity and dielectric behavior and derive meaningful, reaction-specific descriptors.
Combined with our linear mixing approach for the composition, this opens up the search of the entire compositions space.

These similarity metrics with materials properties play a key role in guiding the optimization.
For example, in ORR and HER systems, the similarity to conductivity terms help identify materials with electronic properties suited for fast charge transfer.
Conversely, in OER systems, the similarity to dielectric terms favors materials more aligned with its electronic requirements.
We demonstrated how this process translates qualitative descriptions of materials into quantitative metrics, to filter high performance candidate compositions in large search spaces with high compositional resolution.

The use of text-derived similarities aligns well with the broader goal of electrocatalyst discovery: connecting material compositions and their expected properties to performance metrics in a systematic way.
However, the reliance on embeddings derived solely from textual descriptions also raises questions about the completeness of the data, as some material properties might not be explicitly mentioned in the abstracts.

\subsection{Challenges and Next Steps}

Despite our demonstrated success, our approach has room for improvement:
\begin{itemize}
    \item \textbf{Data Quality:} The quality of the results depends on the quality of the initial text corpus. Missing or biased data can affect the embeddings and the optimization process. We are currently exploring strategies for strategic corpus-building. This will be published in a separate contribution.
    \item \textbf{Balancing Objectives:} Defining the correct optimization goals and weights requires expertise and it is yet unclear how to formalize this process beyond our ``lucky guess''.
\end{itemize}

\section{Conclusions}

Our study outlines an approach for discovering high-performance electrocatalyst materials using automated text mining, Word2Vec models, and Pareto front analysis with high resolution in composition space where experimental data is usually scarce.
Pareto optimization is used to refine large datasets with many candidate compositions into smaller, focused collections of potentially high-performing candidates, tailored to the unique requirements of oxygen reduction (ORR), hydrogen evolution (HER), and oxygen evolution (OER) reactions.
Validated on multiple experimentally-measured datasets, our method reliably identifies compositions with high performance, significantly reducing the candidate pool markedly while maintaining strong property correlations.

\section*{Author Contributions}
\textbf{Lei Zhang:} Conceptualization, Methodology, Software, Validation, Formal analysis, Investigation, Data Curation, Writing - Original Draft, Visualization, Experimentation, Funding acquisition.\\
\textbf{Markus Stricker:} Conceptualization, Resources, Supervision, Writing - Review \& Editing.

\section*{Conflicts of interest}
There are no conflicts to declare.

\section*{Acknowledgements}
Both authors gratefully acknowledge the financial support provided by the China Scholarship Council (CSC, CSC number: 202208360048), which is instrumental in facilitating this research. 
The authors express their gratitude to Lars Banko, Wolfgang Schuhmann, and Alfred Ludwig for their invaluable experimental support and for providing the data used to validate the prediction results. Additionally, the authors acknowledge funding by the Deutsche Forschungsgemeinschaft (DFG, German Research Foundation) through CRC 1625, project number 506711657, subprojects INF, A05.




\bibliography{Zhang_arxiv}
\bibliographystyle{unsrt}

\end{document}